# A discrete memristor model and its application in the Rulkov neuron

Li Jun Liu [1], Du Qu Wei [1]†

[1] College of Electronic Engineering, Guangxi Normal University, Guilin 541004, China

*Abstract*：Continuous-time memristor have been widely used in fields such as chaotic oscillating circuits and neuromorphic computing systems, but research on the application of discrete memristors haven't been noticed yet. In this paper, we designed a new chaotic neuron by applying the discrete model to two-dimensional Rulkov chaotic neuron, and analyzed its dynamical behaviors by extensive experiments involves phase diagram, bifurcation diagram, and spectral entropy complexity algorithm. The experimental results show that the charge of the memristor has an important effect on the system dynamics, delaying the occurrence of bifurcation, and even in the case of full memory ($R_A = 9.4$), leading to the disappearance of the bifurcation thus make the system reach a fixed point. Besides, the increase of the current magnification $k$, can also bring about the increase in discharge frequency of neurons, and a wider and larger range of spectral entropy complexity. The results of our study show the performance of Rulkov chaotic neuron is improved by applying the discrete memristor, and may provide new insights into the mechanism of memory and cognition in the nervous.

**Key words**: discrete memristor; Rulkov neuron; spectral entropy complexity; bifurcation; phase diagram

† E-mail: duquwei@163.com

## 1. Introduction

Current, voltage, charge, and magnetic flux are the four basic variables in the circuit and there is a relationship between two of them. (however, the relationship between electric charge and magnetic flux has never been discovered until Prof. Chua[1] proposed use memristor to describe the relationship between charge and magnetic flux,about 50 years ago. The reason why it is called memristor is because the resistance of memristive is determined by the electric charge flowing through it, so it has the function of memorizing electric charge. Until 2008，Hewlett-Packard [2] William and other talents confirmed the existence of the fourth electronic component memristor for the first time. Because the memristor reaches the nanometer level and is non-volatile, it has been widely used in the field of nonlinear science[3-5].Zhang et al. [6] proposed a chaotic circuit based on a memristor-capacitor, which has abundant chaotic bifurcation paths, various periodic limit cycles and chaotic attractors. [7] studied a new type of simple chaotic circuit based on memristor-capacitor and found 19 different chaotic attractors in the circuit. [8] proposed an unbalanced memristor-synaptic coupling neuron network, revealing several types of collective behavior, including incoherent, coherent, incomplete synchronization, and chimera states. Ma et al. [9][10][11] respectively proposed a new 4D HR model with a threshold flux control memristor (MHR) and a memristive neuron model with adaptive synapses to describes the electromagnetic induction effect. Study the synchronization between memristive Rössler oscillators by reactivating a memristive variable. [12] studied the influence of electromagnetic radiation on the dynamics of spatiotemporal modes in neural networks based on memristors, which are based on nonlinear continuous-time mathematical models[13]. On the other hand, in recent years, discrete neurons have received extensive attention as an effective model for studying neural dynamics. Because it shows high computational efficiency, and the model is simple, reliable, numerically stable, and each iteration has a large time step, it can capture the important dynamics and characteristics of neurons[14-18]. Currently, the well-known discrete-time models include Rulkov model[19-22], Kinouchi and Tragtenberg model[23], Courbage-Nekorkin-Vdovin model[24], etc. Especially, Rulkov chaotic neuron has shown very rich nonlinear dynamic behavior and remarkable biological neuron characteristics, and been widely used in the field of computational neuroscience.[25,26] respectively proposed a discrete integer-order and a fractional-order memristor mathematical models, and apply into the sine chaotic map and the Hénon map. Cao et al.[27]concerns the intermittent evolution routes to the asymptotic regimes in the Rulkov map, successfully predict the evolution path using a three-layer feedforward neural network. Wang et al. [28] discussed the triggering mechanism of the chaotic discharge of the Rulkov neuron model and the mechanism of their chaotic-rest state transition. Hu et al. [29] studied the dynamics such as stability and synchronization of two coupled Rulkov neurons in the presence of electrical synapses and chemical synapses. Tanaka [30] et al. used a mapping-based model to study the firing patterns in neural networks, clarifying the difference between in-phase and anti-phase synchronization patterns and how it exists in the local stability of the invariant subspace. It can be seen that the Rulkov chaotic neuron is a neuron model with simple structure and rich dynamics. Whether the discrete memristor neuron can further improve the performance of the Rulkov chaotic neuron is worth studying.

In this paper, we designed a new chaotic neuron by applying the discrete model to two-dimensional Rulkov chaotic neuron, and analyzed its dynamical behaviors by extensive experiments involves phase diagram, bifurcation diagram, and spectral entropy complexity algorithm. The experimental results show that the charge of the memristor has an important effect on the system dynamics, delaying the occurrence of bifurcation, and even in the case of full memory $(R_A = 9.4)$, leading to the disappearance of the bifurcation thus make the system reach a fixed point. and the value of $R_A$ has an important influence on the chaos of discrete memristors based Rulkov neurons. As the value of $R_A$ increases, the chaotic area of the system gradually shrinks to only bifurcations, and eventually disappears When it increases further. Besides, the increase of the current magnification $k$, can also bring about the increase in discharge frequency of neurons, and a wider and larger range of spectral entropy complexity. The results of this study show that the use of discrete memristors can improve the performance of Rulkov chaotic neurons and may provide new insights for the memory and cognitive mechanisms of the nervous system, as well as for the recognition, diagnosis and treatment of neurological diseases.

The rest of this article is organized as follows. Section 2 introduces the definition of discrete memristor. Section 3 introduces the two-dimensional Rulkov chaotic neuron model based on discrete memristor. Section 4 discusses dynamic analysis of the proposed model from three aspects: bifurcation diagram, time sequence diagram, and general entropy complexity algorithm. Finally, we summarized the research results and looked forward to future research.

## 2. The discrete HP memristor model

Firstly, we consider the classic HP memristor model. What it considers is a semiconductor film with a thickness of D and sandwiched between two metals. This semiconductor film consists of two parts, one part is a small resistance with a high concentration of dopants($R_{ON}$),the remaining part is a large resistance with almost no dopant concentration($R_{OFF}$).Its expression is as follows

$$M(r) = R_{OFF}\left(1 - \frac{\mu_v R_{ON}}{D^2} q(t)\right) \quad (1)$$

respectively. $\mu_v$ is the mobility of doped ions, where $R_A = R_{OFF}$ and $R_B = \frac{R_{OFF} \mu_v R_{ON}}{D^2}$. Then $M(r)$ can be written as

$$M(r) = R_A - R_B q(t) \quad (2)$$

According to Chua, the general continuous-time memristor model mathematical expression is as follows

$$\begin{cases} V(t) = Mq(t)i(t) \\ \dfrac{dq}{dt} = Ki(t) \end{cases} \quad (3)$$

Where $V(t)$ is the voltage of the memristor, and $i(t)$ is the input current, $k$ as a constant can also represent the magnification of the current. The ideal relationship between charge and input current is

$$q(t) = k\int_{-\infty}^{t} i(t)dt \quad (4)$$

The initial state of the memristor is $q(t_0)$, then the $q(t)$ can be written as

$$q(t) = k\int_{-\infty}^{t} i(t)dt = q(t_0) + k\int_{t_0}^{t} i(t)dt \quad (5)$$

$q_n, i_n, v_n$ Respectively represent the value of $q(t), i(t), v(t)$, The discrete memristor equation is as follows

$$\begin{cases} V_n = M(q_n)i_n \\ \nabla q_n = Ki_n \end{cases} \quad (6)$$

where $\nabla q_n = q_n - q_{n-1}, n = 0, 1, 2, \ldots$ in the discrete domain. It can be derived from equation $q_n - q_{n-1} = ki_n$:

$$\begin{aligned} ki_n + ki_{n-1} + k_{n-2} + \ldots + ki_3 + ki_2 + ki_1 + ki_0 \\ = q_n - q_{n-1} + q_{n-1} - q_{n-2} + q_{n-2} \ldots + q_1 - q_0 + q_0 - q_{-1} \\ = q_n - q_{-1} \end{aligned} \quad (8)$$

The expression obtained by shifting the term is

$$q_n = q_{-1} + k\sum_{j=0}^{n} i_j \quad (9)$$

In the discrete domain, $q_{-1}$ can be regarded as the initial electric charge of the memristor, the discrete HP memristor is obtained by

$$V_n = M(r)i_n \quad (10)$$

Then it is derived as

$$V_n = \left[ R_A - R_B(q_0 + K\sum_{j=0}^{n} i_j) \right] i_n \quad (11)$$

### 3. The discrete memristor-based Rulkov model

The Rulkov model is a classic two-dimensional neuron model in computer neuroscience, and has been fully studied since it was proposed in

2003. There are at least three variants of this model, we will refer to the first one as the non-chaotic Rulkov model [31], the second one as the supercritical Rulkov model [32] (which is also non-chaotic),and the third one as the chaotic Rulkov model [33]. We consider the chaotic two-dimensional Rulkov model with the following dynamic equation

$$x(n+1) = \frac{\alpha}{1+x_n^2} + y_n \quad (12)$$

$$y(n+1) = y(n) - \mu(x_n - \sigma) \quad (13)$$

Where $n$ is the discrete time $(n=1,2,...)$, $x_n$ represents the fast variable of neuron transmembrane voltage and $y_n$ represents the slow variable of the gating process. Besides $\alpha, \mu, \sigma$ as system parameters are all variable,where $0 < \mu < 1$. When the system parameters take different values, it shows very rich nonlinear dynamic behavior and significant biological neuron characteristics, the discharge mode is shown in Figure 1, where in Figure 1(a),(b),(c)and(d) correspond to $\alpha = 4, 2, -1, -4$ respectively. Figure 1(a) is periodic bursting, that is, the firing activity of neurons is shown as the alternating state of resting state and chaotic spike discharge state.spike discharge presented in Fig.1b can periodically generates an action potential (Discharge spike) which corresponds to a stable large limit cycle attractor.Figure 1(c) Figure 1(d) respectively show a silent state and a periodic pulse discharge state

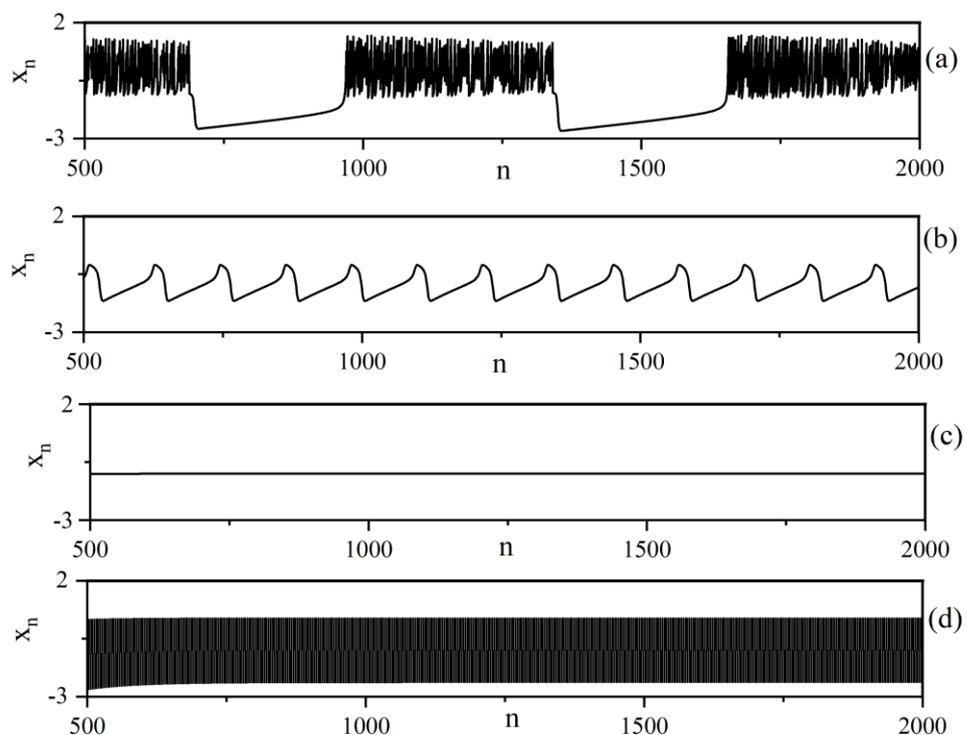

Fig.1 The phase portraits of the fast variable of Rulkov neuron when considering different serval values of $\alpha$,where Figs.(a)-(d)correspond to $\alpha = 4, 2, -1, -4$, respectively.

In order to study the dynamic behavior of the two dimensi-onnal discrete neural system, We apply the the discrete HP memri-stor into the Rulkov chaotic neuron thus propose a discrete two dime-nsional memristive neuron model. Its quation derive as

$$\begin{cases} x(n+1) = \dfrac{\alpha}{1+x_n^2} + \left[R_A - R_B(q_0 + K\sum_{j=0}^{n} i_j)\right] y_n \\ y(n+1) = y(n) - \mu(x_n - \sigma) \end{cases} \quad (14)$$

The use of the memristor leads to an increase in dimensionality. At this time, the system is a three-dimensional memristive nervous system and final three-dimensional equation is as follows

$$\begin{cases} z(n+1) = z_n + y_n \\ x(n+1) = \dfrac{\alpha}{1+x_n^2} + [R_A - R_B(q + kz(n))] y_n \\ y(n+1) = y(n) - \mu(x_n - \sigma) \end{cases} \quad (15)$$

**4. Nonlinear dynamics of the discrete memristor-based Rulkov model**

This section studies the dynamics of the discrete-memristor-based Rulkov neuron model through numerical simulation. The initial state of the system is set to $x_0 = 1$, $y_0 = -2.9$, $z_0 = 0$, the parameters of the neuron are set to $\mu = 0.001$, $\sigma = 0.001$, $R_B = 0.001$, $k = 10$. First, we draw the bifurcation diagram of Rulkov chaotic neurons under different memristor parameters, shown in Figs.2(a)-(f). Fig.2a is the bifurcation of the original neuron ($R_A = 0$), it can be seen that the system bifurcation near $\alpha = 2.9$ and has a wide range of chaotic areas. Figs. (b-f) is the dynamic behavior of discrete memristive Rulkov neurons, corresponding to the parameters $R_A = 0.002, 0.2, 2, 8, 9.4$. From Fig.2b, it can be found that the bi-furcation of the two systems are similar, but the bifurcation threshold of the latter is significantly shifted back, and the bifurcation does not appear until $\alpha = 3.25$, at the same time, the area of chaos is relatively reduced.

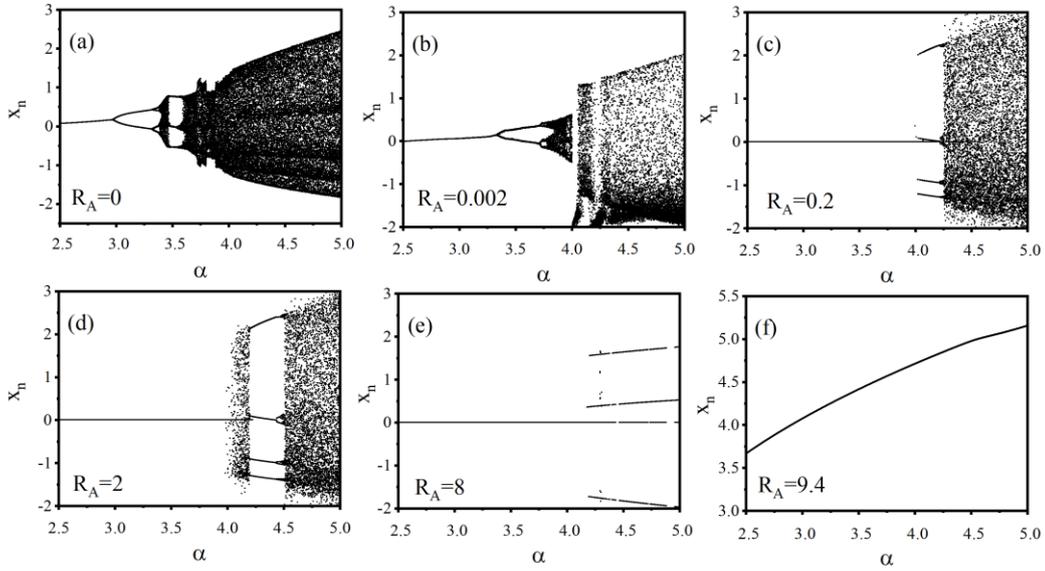

Fig.2 The bifurcation of the neuron with different serval values of $R_A$, where Figs.(a)-(f) correspond to $R_A = 0, 0.002, 0.2, 2, 8, 9.4$, respectively.

with increasing of the value of $R_A$, the tendency of the threshold to move backward becomes more obvious, as can be seen from Figs.2(c-e) which correspond to the thresholds 3.9, 4.1 4.2. especially in Fig.2(f) ($R_A = 9.4$), with the disappeared bifurcation, the neuron is in a resting state. The experimental results show that the resistance of the memristor has an important influence on the system dynamics, delaying the occurrence of bifurcation, and the chaotic region of the system will shrink with the increase of $R_A$ until it finally disappears.

We further verify the effect of current amplification on the firing pattern of neurons, and set the parameters as $\mu = 0.001$, $\sigma = -1$, $\alpha = 3.5$, $R_A = 0.0001$, $R_B = 0.0001$, $q = 0.01$. The results are shown in Figs.3(a)-(d), which correspond to $k = 0.2, 0.3, 0.5, 1$. From Figure 3 we can observe that as the current magnification increases, the neuron cluster firing becomes more frequent. which indicates that an increase in current amplification $k$ can induce and enhance neuronal activity.

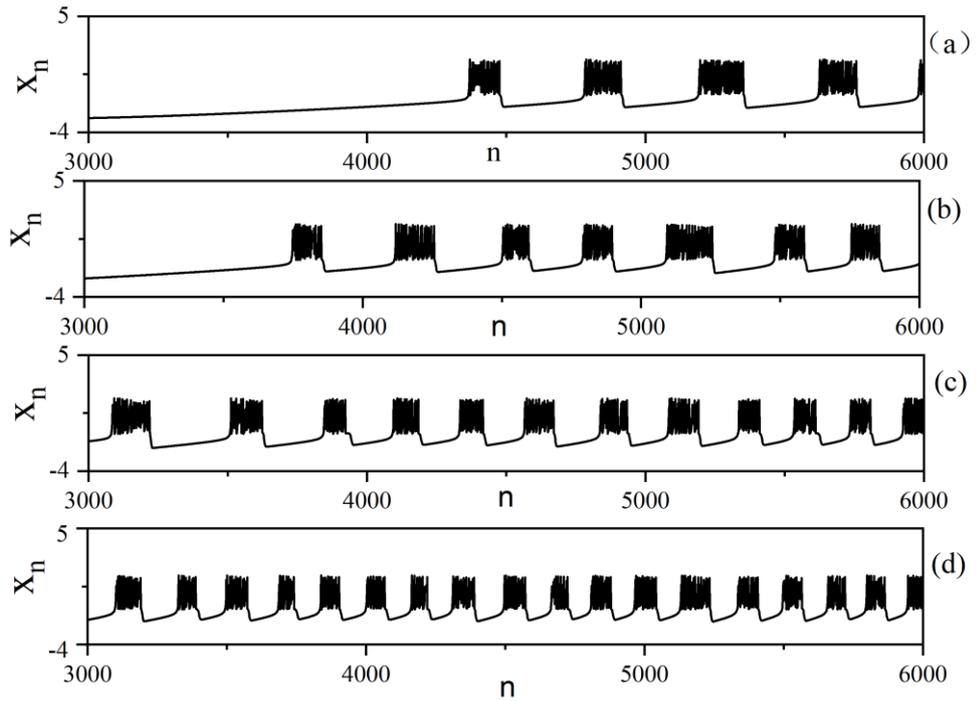

Fig.3 The phase portraits of the fast variable of the discrete memristive Rulkov neuron considering different serval values of $k$, where Figs.(a)-(d) correspond to $k = 0.2, 0.3, 0.5, 1$, respectively.

Finally, study the complexity of the discrete memristive Rulkov neuron model. Proven by previous work the chaotic activity characteristics of EEG(electroencephalo-gram) are closely related to the functional state of the brain, in the normal state of the brain, the dimensionality, Lyapunov index, complexity and other indicators of the chaotic activity of the brain are relatively high; while the above-mentioned chaotic indicators will decrease if the brain is in the pathological state of impaired brain function.Among them, complexity is of great significance to the normal activities of brain function which is why we mainly pay attention to it.Complexity is an indicator used to measure the closeness of a random sequence to a chaotic sequence. In short, the greater the complexity, the more random the time series.In this paper,we use spectral entropy complexity (SE) algorithm [34] to calculate the complexity of these two systems, and set the number of samples in the time series as 30000. Fig. 4(a) and (b) respectively show the complexity results of the original Rulkov chaotic neuron and the discrete memristive Rulkov chaotic neuron in sequence. From the figure, we can clearly see that the complex light-colored area of the original Rulkov neuron occupies a wider range，while the dark-colored area is mainly concentrated in a small part where $\alpha \in (6,8]$.On the other hand, the discrete memristive Rulkov neuron with high complexity occupy wider range in the Fig.4(b) where $\alpha \in (4,10]$. With the high complexity, discrete memristive Rulkov chaotic neuron is more adaptable to internal and external environments, more reliable, easier to accept and process external information, and more adaptable to extracellular factors such as neurotransmitters, temperature, etc. According to [35-37], the complexity of EEG activity increases sequentially in three states: coma, deep sleep, and awake. Therefore, discrete memristive Rulkov chaotic neurons have better prospects for medical applications, and have potential significance for the understanding, diagnosis and treatment of neurological diseases.

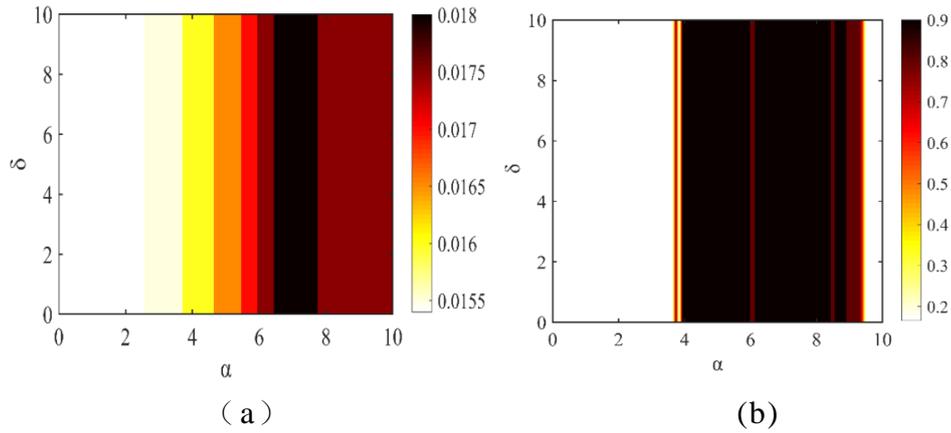

（a）　　　　　　　　　　　　（b）

**5. conclusion**

Because discrete memristors are more suitable for discrete chaotic systems and digital circuits, we apply discrete HP memristors into two-dimensional Rulkov chaotic neurons, named discrete memristive Rulkov chaotic neurons. Its dynamical behaviors are analyzed by phase diagram, bifurcation diagram, and spectral entropy complexity algorithm. Experimental results show, increasing the low resistance $R_A$ value of the memristor with a high concentration of dopants delays the occurrence of bifurcation. As the resistance value increases, the chaotic area of the neuron decreases, leading to the disappearance of the bifurcation, thus make the neuron stay in a resting state. The current magnification $k$ also has an important influence on neuron activity. The increase of the current magnification can bring about a significant increase in the frequency of neuron discharge. Compared with the original system, the spectral entropy complexity of this system has a wider range of complexity and higher numerical value, which is closer to a random sequence and can be expected to provide a new method for the recognition, diagnosis and treatment of neurological diseases. The results of this study show that the use of discrete memristors can improve the performance of Rulkov chaotic neurons and may provide new insights into the memory and cognitive mechanisms of the nervous system.